\documentclass[doublecol]{epl2}
\usepackage{amsmath}
\usepackage{amssymb}
\usepackage{latexsym}
\usepackage{sistyle}

\title{Steady state fluctuation relations for systems driven by an external random force}
\shorttitle{Fluctuation relations for systems driven by an external random force} %Insert here a short version of the title if it exceeds 70 characters

\author{J. R. Gomez-Solano\inst{1} \and L. Bellon\inst{1} \and
  A. Petrosyan\inst{1} \and S. Ciliberto\inst{1}}
\shortauthor{J. R. Gomez-Solano \etal}

\institute{
  \inst{1} Universit\'e de Lyon, Laboratoire de Physique, Ecole Normale Sup\'erieure de Lyon, CNRS -
  46, All\'ee d'Italie, 69364 Lyon CEDEX 07, France\\
}
\pacs{05.40.-a}{Fluctuation phenomena, random processes, noise, and Brownian motion}
\abstract{We experimentally study the fluctuations of the work
done by an external Gaussian random force on two different
stochastic systems coupled to a thermal
bath: a colloidal particle in an optical trap and an atomic force
microscopy cantilever. We determine the corresponding probability
density functions for different random forcing amplitudes ranging
from a small fraction to several times the amplitude of the
thermal noise. In both systems for sufficiently
weak forcing amplitudes the work fluctuations satisfy the usual
steady state fluctuation theorem. As the forcing amplitude drives the
system far from equilibrium, deviations of the fluctuation theorem
increase monotonically. The deviations can be recasted to a single master curve
which only depends on the kind of stochastic external
force.}

\begin{document}

\maketitle

\section{Introduction}

Fluctuation relations are a very important theoretical result for
the description of non-equilibrium
microscopic systems since they quantify the statistical properties of
fluctuating energy exchanges under rather general conditions \cite{kurchan}.
In particular, the so-called fluctuation theorem (FT) \cite{evans,gallavotti}
quantifies the asymmetry of the distribution of positive and negative
fluctuations of a given time-integrated quantity (injected work, entropy production, etc.).
For a system in contact with a thermostat at temperature $T$ and
driven by an external force in a non-equilibrium steady state,
the FT states that the ratio of the probability of finding a positive fluctuation
with respect to that of the corresponding negative value for the work $W_{\tau}$
done by the force during a time interval $\tau$ satisfies
\begin{equation}
\label{eq.SSFT}
     \ln \frac{P(W_{\tau}=W)}{P(W_{\tau}=-W)} \rightarrow  \frac{W}{k_B T}, \, \tau_c \ll \tau,
\end{equation}
where $\tau_c$ is the longest characteristic relaxation time of
the system. Equation~(\ref{eq.SSFT}) has been tested in several
experiments such as fluidized granular media \cite{feitosa}, a
colloidal particle dragged by an optical trap \cite{wang},
electrical circuits \cite{garnier}, mechanical harmonic
oscillators \cite{joubaud} and a colloidal particle near the
stochastic resonance \cite{jop}. New fluctuation relations have
been proposed as well for the entropy production \cite{seifert} or
by considering modifications of the statistical properties of the
thermal bath \cite{zamponi,anomalousFR,baule}. In all of these
examples the force which drives the system out of equilibrium is
inherently deterministic. However, it has been recently argued
that the nature (deterministic or stochastic) of the forcing can
play an important role in the distribution of the injected work
leading to possible deviations from the relation (\ref{eq.SSFT})
for large fluctuations ($W_{\tau}/\langle W_{\tau}\rangle > 1)$.
Indeed, it has been found in experiments and simulations such as a
Brownian particle in a Gaussian white \cite{farago} and colored
\cite{farago2} noise bath, turbulent thermal convection
\cite{shang}, wave turbulence \cite{efalcon}, a vibrating metalic
plate \cite{cadot}, an RC electronic circuit \cite{cfalcon} and a
gravitational wave detector \cite{bonaldi} that the probability
density functions of the work done by a stochastic force are not
Gaussian but asymmetric with two exponential tails leading to
violations of the FT in the form of effective temperatures or
nonlinear relations between the left and the rigth hand side of
eq.~(\ref{eq.SSFT}). It is important to remark that in the systems
previously cited the steady state FT is violated because in such a
case the external random force acts itself as a kind of thermal
bath. One question which naturally arises is what the work
fluctuation relations will become when in addition to the
external random forcing a true thermalization process is allowed.
In this situation there are two sources of work fluctuations: the
external force and the thermal bath. As pointed out in
\cite{baule,cadot}, one is interested in the distribution of the work
fluctuations done by the external random force in presence of a
thermostat and the conditions under which the FT could be valid.

In the present work we address these questions in two experimental
systems: a Brownian particle in an optical trap and a
micro-cantilever used for atomic force microscopy (AFM). Both are
in contact with a thermal bath and driven out of equilibrium by an external random
force whose amplitude is tuned from a small fraction to several
times the amplitude of the intrinsic thermal fluctuations exerted
by the thermostat.

\begin{figure}[h]
\begin{center}$
\begin{array}{cc}
\includegraphics[width=1.35in]{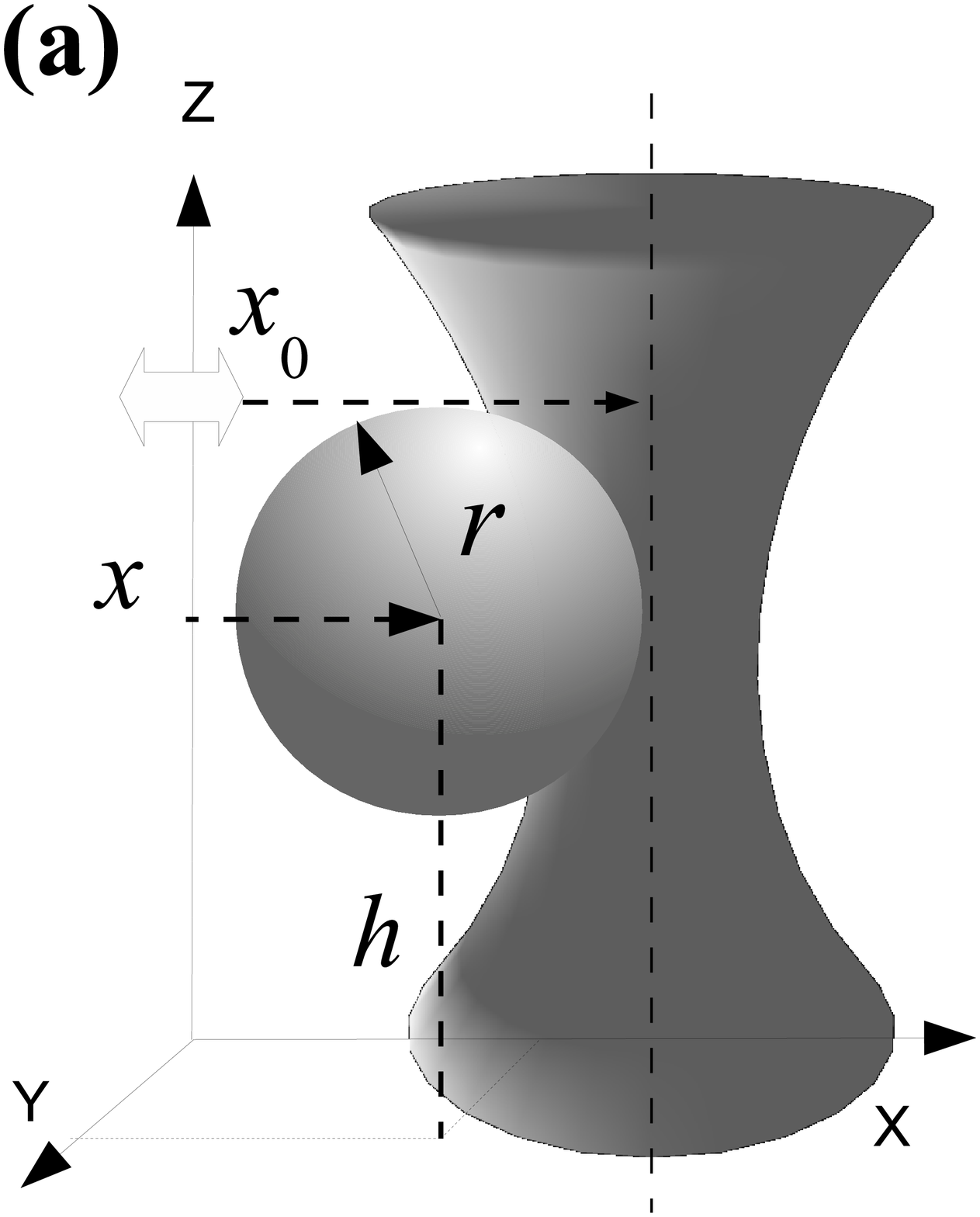} &
\includegraphics[width=1.85in]{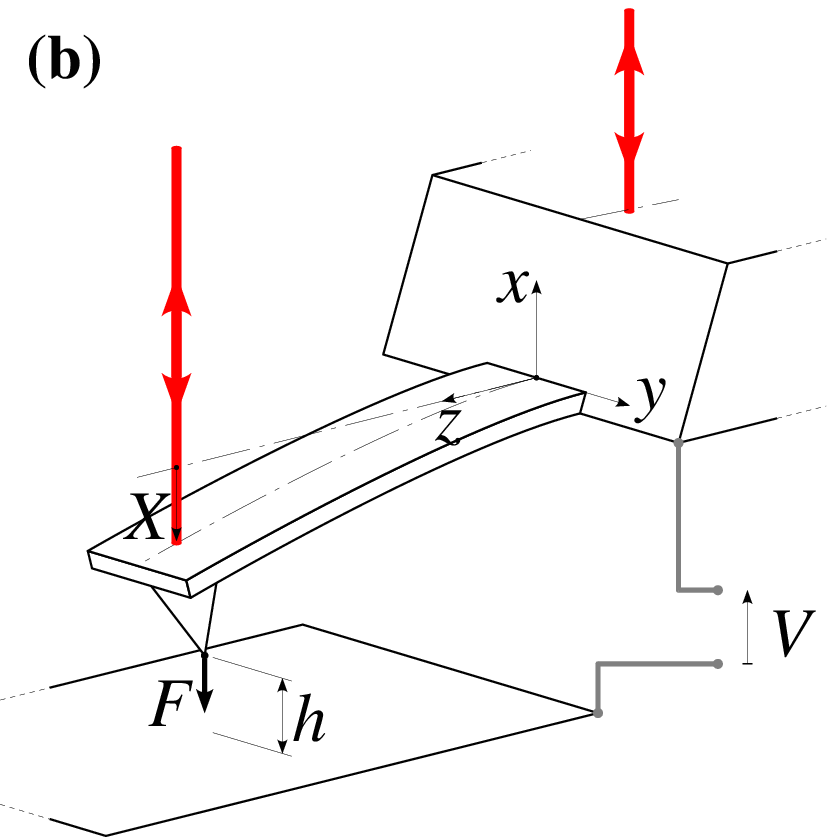}
\end{array}$
\end{center}
\caption{a) Colloidal particle in the optical trap with modulated position.
b) AFM  cantilever close to a metallic surface. See text for explanation. }
\label{fig:setup}
\end{figure}

\section{Colloidal particle in an optical trap}
\begin{figure}%[htp]
 \centering{\includegraphics[width=.49\textwidth]{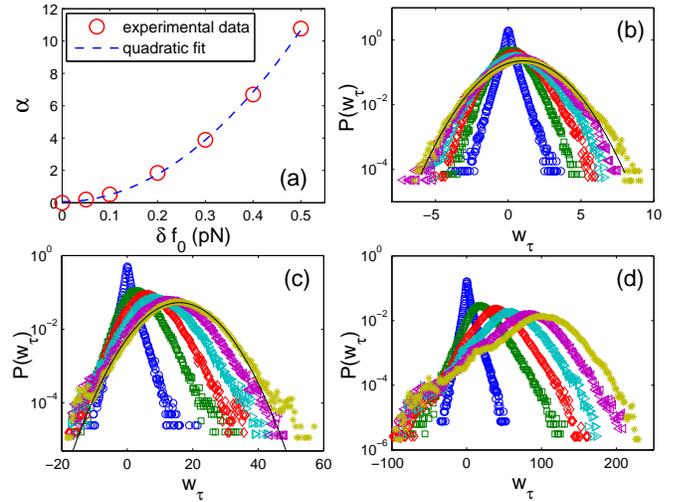}}
     \caption{(a) Dependence of the parameter $\alpha$ on the standard deviation
    of the Gaussian exponentially correlated external force $f_0$ acting on
the colloidal particle. (b) Probability
density functions of the work $w_{\tau}$ for $\alpha = 0.20$; (c) $\alpha = 3.89$;
and (d) $\alpha = 10.77$. The symbols correspond to integration times $\tau =5$ ms
($\circ$), 55 ms ($\Box$), 105 ms ($\Diamond$), 155 ms ($\triangleleft$),
205 ms($\triangleright$) and 255 ms ($\ast$). The solid black lines in (b) and (c)
are Gaussian fits.}
  \label{fig:PDFwork_OPT}
\end{figure}

The first system we study consists on a spherical silica bead of
radius $r = \SI{1} {\mu m}$ immersed in ultrapure water which acts as a
thermal bath. The experiment is performed at a room temperature of
$27 \pm 0.5^{\circ}$C at which the dynamic viscosity of water is
$\eta = (8.52 \mp 0.10) \times 10^{-4}$ Pa s. The motion of the
particle is confined by an optical trap which is created by
tightly focusing a Nd:YAG laser beam ($\lambda = \SI{1064}{nm}$) by
means of a high numerical aperture objective (63$\times$, NA =
1.4). The trap stiffness is fixed at a constant value of $k = \SI{5.4} {pN/\mu m}$.   
The particle is kept at $h \approx \SI{10} {\mu m}$ above the lower cell surface
to avoid hydrodynamic interactions with the walls. Figure~\ref{fig:setup}(a)  sketches the
configuration of the bead in the optical trap. An external random
force is applied to the particle by modulating the position of the
trap $x_0(t)$ using an acousto-optic deflector, along a fixed
direction {\bf x} on the plane perpendicular to the beam propagation (+{\bf z}). The
modulation corresponds to a Gaussian Ornstein-Uhlenbeck noise of
mean $\langle x_0(t) \rangle=0$ and covariance $\langle x_0(s)
x_0(t) \rangle = A \exp(-|t - s| / \tau_0)$. The correlation time
of the modulation is set to $\tau_0 = \SI {25}{ms}$ whereas the value of
its amplitude $A$ is tuned to control the driving intensity. We
determine the particle barycenter $(x,y)$ by image analysis using
a high speed camera at a sampling rate of $\SI{1} {kHz}$ with an accuracy
better than $\SI{10} {nm}$. See ref.~\cite{laponite} for more details about
the experimental apparatus. The attractive force exerted by the
optical trap on the bead at time $t$ along {\bf x} is given by 
$-k(x(t) - x_0(t))$. Hence, for the
experimentally accessible timescales the dynamics of the
coordinate $x$  is described  by the overdamped Langevin equation
\begin{equation}
\label{eq.1}
    \gamma \dot{x} = -kx + \zeta_T + f_0.
\end{equation}
In eq.~(\ref{eq.1}) $\gamma = 6\pi r \eta$ is the viscous drag coefficient,
$\zeta_T$ is a Gaussian white noise ($\langle \zeta_T \rangle=0$,
$\langle \zeta_T(s) \zeta_T(t)\rangle = 2k_B T \gamma \delta(t-s)$)
which mimics the collisions of the thermal bath particles with the colloidal bead and
$f_0(t) = k x_0(t)$ plays the role of the external stochastic force.
The standard deviation  $\delta f_0$ of $f_0$ is chosen as the main control parameter of the system.
Besides the correlation time $\tau_0$ of $f_0$ there is a second characteristic timescale
in the dynamics of eq.~(\ref{eq.1}): the viscous relaxation time in the optical
trap $\tau_{\gamma} = \gamma/k = \SI{3} {ms} < \tau_0$.
In order to quantify the relative strength of the external force with respect
to the thermal fluctuations, we introduce a dimensionless parameter which measures
the distance from equilibrium
\begin{equation}
\label{eq.2}
    \alpha = \frac{\langle x^2 \rangle}{\langle x^2 \rangle_{eq}} - 1,
\end{equation}
where $\langle x^2 \rangle$ is the variance of $x$ in the presence of $f_0>0$ whereas
$\langle x^2 \rangle_{eq}$ is the corresponding variance at equilibrium ($f_0=0$).
The dependence of $\alpha$ on $\delta f_0$ is quadratic, as shown in
fig.~\ref{fig:PDFwork_OPT}(a). This quadratic dependence is a consequence of the
linear response of the system to the external forcing described by the 
linear Langevin eq.~(\ref{eq.1}).

The work done by the external random force on the colloidal
particle (in $k_B T$ units) is
\begin{equation}
\label{eq.workrandom}
    w_{\tau}= \frac{1}{k_B T}\int_t^{t+\tau} \dot{x}(t')f_0(t')\mathrm{d}t'.
\end{equation}
%are the potential energy variation, the work done by the thermal bath,
%the heat transferred to the bath due to viscous dissipation and the work
%done by the external force, respectively, divided by the temperature.
Thus, by measuring simultaneously the time evolution of the
barycenter position of the particle and the driving force we are
able to compute directly the work injected into the system by the
driving. In figs.~\ref{fig:PDFwork_OPT}(b)-(d) we show the
probability density functions (PDF) of $w_{\tau}$ for different
values of $\tau$ and $\alpha$. We observe that for a fixed value
of $\alpha$, the PDFs have asymmetric exponential tails at short
integration times and they become smoother as the value of $\tau$
increases. For $\alpha = 0.20$ they approach a
Gaussian profile (fig.~\ref{fig:PDFwork_OPT}(b)) whereas
asymmetric non-Gaussian tails remain for increasing values of
$\alpha$. As shown in
figs.~\ref{fig:PDFwork_OPT}(c)-(d), the asymmetry of these tails
becomes very pronounced for large $\alpha
> 1$ even for integration times as long as $\tau = \SI{250}{ms}=10
\tau_0$, where we have taken $\tau_0$ because it is the largest
correlation time of the dynamics. As pointed out in
\cite{cfalcon}, the deviations of the linear relation of
eq.~(\ref{eq.SSFT}) (with respect to $w_{\tau}$) can occur for
extreme values of the work fluctuations located on these tails.

\begin{figure}%[htp]
 \centering{\includegraphics[width=.49\textwidth]{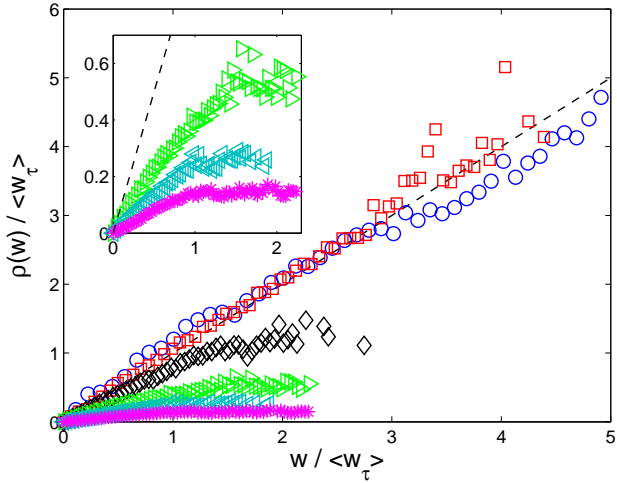}}
     \caption{Asymmetry function of the PDF of the work done
     by the external force on the colloidal bead computed at $\tau = 10\tau_0$ for
     different values of the parameter $\alpha$: $0.20 (\circ), 0.51 (\Box), 1.84 (\Diamond),
     3.89 (\triangleright), 6.69 (\triangleleft), 10.77 (\ast)$. The dashed line represents
     the prediction of the fluctuation theorem $\rho(w)=w$. Inset: Expanded view
     for $\alpha \ge 3.89$.}
  \label{fig:FT_OPT}
\end{figure}

We define the asymmetry function of the PDF $P$ as
\begin{equation}
\label{eq.asym}
    \rho(w) = \lim_{\frac{\tau}{\tau_c} \rightarrow \infty}
        \ln \frac{P(w_{\tau}=w)}{P(w_{\tau}=-w)},
\end{equation}
so that eq.~(\ref{eq.SSFT}) reads
\begin{equation}
\label{eq.SSFT1}
    \rho(w)=w.
\end{equation}
From the experimental PDFs of $w_{\tau}$ we compute $\rho(w)$ as
the logarithm in eq.~(\ref{eq.asym}) for integration times $\tau =
10 \tau_0$. We checked that for this value the limit of
eq.~(\ref{eq.asym}) has been attained. Figure~\ref{fig:FT_OPT} shows
the profile of the asymmetry functions for different values of
$\alpha$. We notice that for sufficiently small values ($\alpha
=0.20, 0.51 < 1$), the FT given by eq.~(\ref{eq.SSFT1}) is
verified by the experimental data. To our knowledge, 
this is the first time that the FT holds for 
a random force without introducing any prefactor in the 
linear relation of eq.~(\ref{eq.SSFT1}). 
It is important to point out that any
deviation from the linear relation of eq.~(\ref{eq.SSFT1}) for
extreme fluctuations is unlikely since we probed values as large
as $w_{\tau}/\langle w_{\tau} \rangle \sim 5$. 
Indeed it is argued \cite{baule,farago,farago2,cfalcon}, that, for
strongly dissipative systems driven by a random force, the
deviations from FT may occur around $w_{\tau}/\langle w_{\tau}
\rangle \sim 1$. Furthermore in the present case the validity of
the FT for weak driving amplitudes $\alpha < 1$ is consistent with
the fact that for integration times $\tau > 25$ ms, the ratio
$\rho(w) /w$ has converged to its asymptotic value $1$ for all
measurable $w$. Note that this convergence to the FT prediction is
quite similar to that measured in system driven out of equilibrium 
by deterministic forces \cite{garnier,joubaud,jop}. 
For instance in the case of a harmonic oscillator driven by a sinusoidal
external force the asymptotic value of $\rho(w) /w$ is reached for
integration times larger than the forcing period \cite{joubaud}.

In contrast, deviations from eq.~(\ref{eq.SSFT1}) 
are expected to occur for $1 < \alpha$ because the
fluctuations of injected energy produced by the external random
force become larger than those injected by the thermal bath.
Indeed fig.~\ref{fig:FT_OPT} shows that for values above $\alpha =
1.84$, eq.~(\ref{eq.SSFT1}) is not verified any more but $\rho$
becomes a nonlinear function of $w_{\tau}$. For small values of
$w_{\tau}$ it is linear with a slope which decreases as the
driving amplitude increases whereas there is a crossover to a
slower dependence around $w_{\tau}/\langle w_{\tau} \rangle \sim
1$, a qualitatively similar behavior to those reported in
\cite{farago,efalcon,cfalcon,cadot,bonaldi}. We finish this
section by emphasizing that we have clearly found that for an
experimental system whose dynamics correspond to a first order
Langevin equation subjected to both thermal and external noises,
the FT can be satisfied or not depending on the relative strength
of the external driving. The details about how this deviations
arise and the convergence to generic work fluctuation relations
will be given further. We first analyze the experiment on the AFM.

\section{AFM cantilever}

%\begin{figure}%[htp]
 %\centering{\includegraphics[width=.35\textwidth]{cantilever.eps}}
 %    \caption{
%Sketch of the AFM setup. The deflection $X$ of the cantilever 
%is measured by differential interferometry between the reference beam, 
%reflecting on the chip near the clamping extremity, and the sensing beam, 
%reflecting on the free end of the cantilever. When a voltage $V$ is 
%applied between the conductive tip and the surface, an attractive electrostatic 
%force $F\propto V^{2}$ acts on the cantilever. The surface 
%being close enough (tip sample distance $h\lesssim\SI{10}{\mu m}$), 
%the force is mainly applied to the tip.}
 % \label{fig:AFMsetup}
%\end{figure}

\begin{figure}%[htp]
 \centering{\includegraphics[width=.49\textwidth]{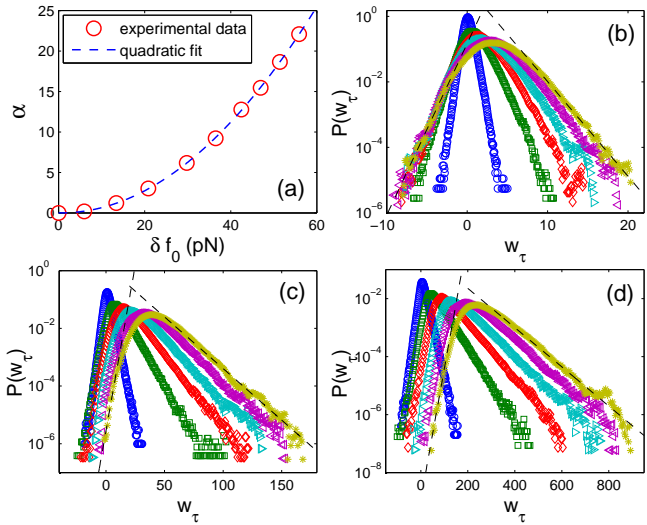}}
     \caption{(a) Dependence of the parameter $\alpha$ on the standard deviation
    of the Gaussian white external force $f_0$ acting on the cantilever. (b) Probability
density functions of the work $w_{\tau}$ for $\alpha = 0.19$; (c) $\alpha = 3.03$;
and (d) $\alpha = 18.66$. The symbols correspond to integration times $\tau = 97$ $\mu$s
($\circ$), 1.074 $\mu$s ($\Box$), 2.051 ms ($\Diamond$), 3.027 ms ($\triangleleft$),
4.004 ms($\triangleright$) and 4.981 ms ($\ast$). The black dashed lines in (b)-(d)
represent the exponential fits of the corresponding tails.}
  \label{fig:PDFwork_AFM}
\end{figure}

A second example of a system for which thermal fluctuations are
non-negligible in the energy injection process at equilibrium is
the dynamics of the free end of a rectangular micro-cantilever
used in AFM measurements. The cantilever is a mechanical
clamped-free beam, which can be bended by an external force $F$
and is thermalized with the surrounding air.
The experiment is sketched
in fig.~\ref{fig:setup}(b).

We use conductive cantilevers from Nanoworld (PPP-CONTPt). 
They exhibit a nominal rectangular geometry: $\SI{450}{\mu m}$ long, 
$\SI{50}{\mu m}$ wide and $\SI{2}{\mu m}$ thick, with a $\SI{25}{nm}$ 
PtIr$_{5}$ conductive layer on both sides. The deflection is measured 
with a home made interferometric deflection sensor \cite{Paolino:afm}, 
inspired by the original design of Schonenberger \cite{1989Schonenberger} 
with a quadrature phase detection technique \cite{2002Bellon}: the interference 
between the reference laser beam reflecting on the chip of the cantilever 
and the sensing beam on the free end of the cantilever gives a direct 
measurement of the deflection $X$.
%A first advantage of this technique is that it offers a calibrated 
%measurement of deflection, without conversion factor from angle to 
%displacement, as in the standard optical lever technique common in AFM. A second advantage of
Our detection system has a very low intrinsic noise, as low as $\SI{4}{pm}$ rms 
in the $\SI{100}{kHz}$ bandwidth we are probing\cite{Paolino:afm,Paolino-2009}.

From the power spectrum of the deflection fluctuations of the free end at 
equilibrium ($F = 0$) we verify that the cantilever dynamics can be 
reasonably modeled as a stochastic harmonic oscillator with viscous 
dissipation\cite{Paolino-2009,Bellon-2008}. Hence, in the presence of the external force the dynamics of the vertical coordinate $X$ of the free end is described by the second order Langevin equation
\begin{equation}
\label{eq.harmonic0}
    m\ddot{X} + \gamma \dot{X} = -kX + \zeta_T + F,
\end{equation}
where $m$ is the effective mass, $\gamma$ the viscous drag
coefficient, $k$ the stiffness associated to the elastic force on
the cantilever and $\zeta_T$ models the thermal fluctuations. $m$, 
$\gamma$ and $k$ can be calibrated at zero forcing using
fluctuation dissipation theorem, relating the observed power
spectrum of $X$ to the harmonic oscillator model: in our
experiment we measure $m = \SI{2.75E-11}{kg}$, $\gamma =
\SI{4.35E-8}{kg/s}$ and $k=\SI{8.05E-2}{N/m}$. The
amplitude of the equilibrium thermal fluctuations of the tip
position (i.e. $\sqrt{\langle x^2\rangle_{eq}}=\sqrt{k_BT/k}\simeq 2 \ 10^{-10}$m)
is two orders of magnitude larger then the detection noise ({\it i.e}.
$\SI{4}{pm}$ rms). The signal to noise ratio is even better when the
system is driven by an external force $F$. The characteristic
timescales of the deflection dynamics are the resonance period of
the harmonic oscillator $\tau_k = 2\pi \sqrt(m/k) = \SI{116}{\mu
s}$ and the viscous relaxation time $\tau_{\gamma} = m / \gamma =
\SI{632}{\mu s}$, which is the longest correlation time.

When a voltage $V$ is applied between the conductive cantilever and
a metallic surface brought close to the tip ($h\sim\SI{10}{\mu m}$ apart),
an electrostatic interaction is created. The system behaves as
a capacitor with stored energy $E_{c}=\frac{1}{2} C(X) V^2$, with $C$
the capacitance of the cantilever-tip/surface system. Hence, the
interaction between the cantilever and the opposite charged surface
gives rise to an attractive external force $F = -\partial_{X}E_{c}=-aV^2$
on the free end, with $a=\partial_{X}C/2$. If we apply a static voltage
$\overline{V}$, the force $F$ can be deduced from the stationary solution
of eq.~(\ref{eq.harmonic0}): $k \overline{X} = -a \overline{V}^{2}$, where
 $\overline{X}$ is the mean measured deflection. $k$ being already calibrated,
  we validate this quadratic dependence\footnote{The quadratic
   dependance is valid only after taking care to compensate for the
    contact potential between the tip and the sample, which gives a
     small correction of the order of a few tens of mV.} of forcing in $V$ and measure  $a=\SI{1.49E-11}{N/V^2}$.

As the electrostatic force $F$ is only attractive, its mean value cannot
be chosen to be 0. We thus generated a driving voltage $V$ designed to
create a Gaussian white noise forcing $f_{0}$ around an offset $\overline{F}$:
 $F=\overline{F}+f_{0}$. The variance $\delta f_0$ of $f_{0}$ is the main
 control parameter of the system. In the absence of fluctuations
 $\zeta_T$ and $f_{0}$, eq.~(\ref{eq.harmonic}) has
 the stationary solution $\overline{X}=\overline{F}/k$.
  This solution corresponds to the mean position attained
  by the free end in the presence of the zero mean fluctuating forces.
   Hence, we focus on the dynamics of the fluctuations $x=X-\overline{X}$
    around $\overline{X}$ which are described by the equation
\begin{equation}
\label{eq.harmonic}
    m\ddot{x} + \gamma \dot{x} = -kx + \zeta_T + f_{0}.
\end{equation}

\begin{figure}%[htp]
 \centering{\includegraphics[width=.49\textwidth]{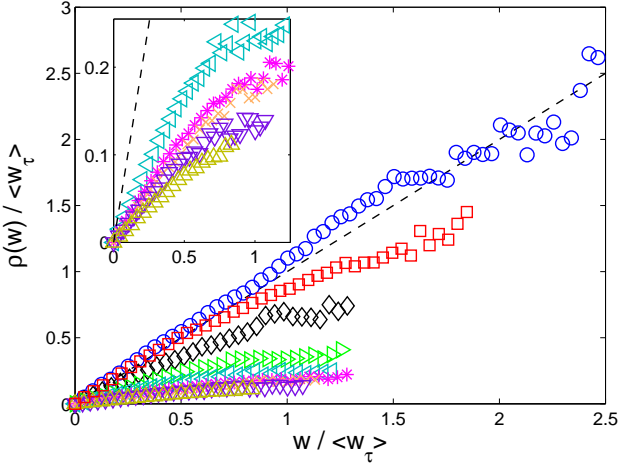}}
     \caption{Asymmetry function of the probability density function of the work done
     by the external force on the AFM cantilever computed at $\tau = 4\tau_{\gamma}$ for
     different values of the parameter $\alpha$: $0.19 (\circ), 1.21 (\Box), 3.03 (\Diamond),
     6.18 (\triangleright), 9.22 (\triangleleft), 12.77 (\ast)$, $15.46 (\times),
     18.66 (\bigtriangledown), 22.10 (\triangle)$. The dashed line corresponds to
     the prediction of the fluctuation theorem $\rho(w)=w$.
     Inset: Expanded view for $\alpha \ge 9.22$.}
  \label{fig:FT_AFM}
\end{figure}

Figure \ref{fig:PDFwork_AFM}(a) shows the dependence between the parameter 
$\alpha$ defined in eq.~(\ref{eq.2}) for the stochastic variable $x$
and the control parameter $\delta f_0$.
We find that this dependence is quadatric verifying the linearity
of the stochastic dynamics of the free end of the cantilever.
On the other hand, the work done by the external random force
during an integration time $\tau$ is computed from
eq.~(\ref{eq.workrandom}). The corresponding PDFs are shown in
figs.~\ref{fig:PDFwork_AFM}(b)-(d). Unlike the
colloidal particle, the PDFs do not converge to a
Gaussian distribution but to a profile with asymmetric exponential
tails even for the smallest driving amplitude ($\alpha =
0.19$) and for integration times as long as $\tau = 8
\tau_{\gamma}$, as shown in figs.~\ref{fig:PDFwork_AFM}(b)-(d).
Surprisingly, when computing the asymmetry function for 
$\alpha = 0.19 < 1$ and $\tau =
4\tau_{\gamma}$ the steady state FT of eq.~(\ref{eq.SSFT1}) is
perfectly verified, as shown in fig.~\ref{fig:FT_AFM}. Work fluctuations
as large as 2.5 times their mean value located on the exponential
tails are probed and hence deviations from FT are unlikely for the
same reasons discussed for the case of the Brownian particle.

In fig.~\ref{fig:FT_AFM} wee see that for $1.21 \le \alpha$, 
the deviations from eq.~(\ref{eq.SSFT1}) appear
as a nonlinear relation with a linear part for small fluctuations
whose slope decreases as $\alpha$ increases and a crossover for
larger fluctuations, qualitatively similar to the behavior
observed for the colloidal particle, as shown clearly in the inset of
fig.~\ref{fig:FT_AFM}. In the following we
discuss the properties of these deviations as the energy injection
process becomes dominated by the external force.

\section{Fluctuation relations far from equilibrium}
We address now the question of how the deviations from eq.~(\ref{eq.SSFT1})
arise as the external stochastic force drives the system far from equilibrium.
As shown previously, for $1 \lesssim \alpha$, the forcing amplitude is strong enough to
destroy the conditions for the validity of the FT for $w_{\tau}$. We note that there
are two well defined limit regimes depending on the driving amplitude:
one occuring at small values of $\alpha$ for which the steady state FT is valid, and the
limit $\alpha \gg 1$ for which the the role of the thermal bath must be negligible in the energy injection
process, which must be completely dominated by the external stochastic force. In order to investigate
whether the transition between these two regimes is abrupt or not, we proceed by
noting that for the latter the stochastic force term $\zeta_T$ in
eqs.~(\ref{eq.1}) and (\ref{eq.harmonic}) will be negligible compared to $f_0$.
This implies that the resulting statistical time-integrated properties of the corresponding
non-equilibrium steady state will be invariant under a normalization of the timescales
and the temperature of the system. In particular, the resulting fluctuation relations
for $w_{\tau}$ must lead to a master curve for the asymmetry function
in the far from equilibrium limit $\alpha \gg 1$. The information about the
transition of the fluctuation relations to this regime is given by the convergence
to the master curve.

\begin{figure}%[htp]
 \centering{\includegraphics[width=.49\textwidth]{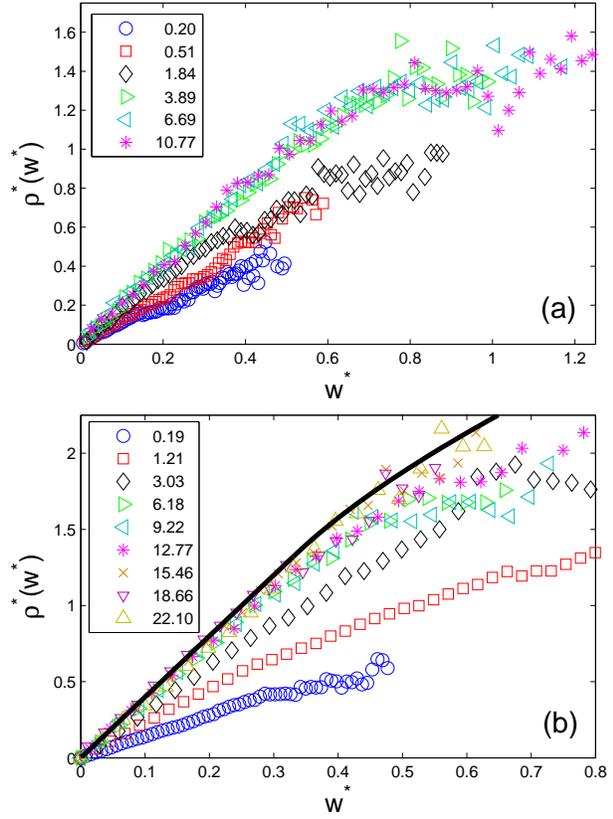}}
     \caption{(a) Asymmetry function of the PDF of the normalized work done by the Gaussian
     Ornstein-Uhlenbeck force on the colloidal particle for different values of the parameter $\alpha$.
     (b) Asymmetry function of the PDF of the normalized work done by the Gaussian white force
     on the cantilever for different values of the parameter $\alpha$. The thick solid line
     represents the analytical expression given by eq.~(\ref{asymptotic_asymmetry}).}
  \label{fig:FTnormalized}
\end{figure}

We introduce the normalized work $w^{*}_{\tau}$ as
\begin{equation}
\label{eq.normalization}
    w^{*}_{\tau}=\frac{\tau_c}{\tau}\frac{w_{\tau}}{1+\alpha}.
\end{equation}
The physical idea behind this normalization is that for $\alpha
\gg 1$, the thermal bath alone works as a heat reservoir for
viscous dissipation  whereas its coupling with the external
forcing plays the role of a non-equilibrium thermal bath at an
effective temperature $k \langle x^2 \rangle/k_B = (1+\alpha)T
\approx \alpha T$. The prefactor $\tau_c / \tau$ is introduced in
such a way that $w^{*}_{\tau}$ represents the average normalized
work done during the largest correlation time of the system.
Accordingly, the asymmetry function must be redefined as
\begin{equation}
\label{eq.normalization_symfunct}
    \rho^{*}(w^{*})=\lim_{\tau/\tau_c\rightarrow \infty}
\frac{\tau_c}{\tau}\ln \frac{P(w^{*}_{\tau} = w^{*})}{P(w^{*}_{\tau} =- w^{*})}.
\end{equation}

Figure.~\ref{fig:FTnormalized}(a) shows the asymmetry function
$\rho^{*}$ for the normalized work $w^{*}_{\tau}$ on the colloidal
particle at large values of $\alpha$ for which
eq.~(\ref{eq.SSFT1}) is violated. The timescale $\tau_c$ in the
computation of (\ref{eq.normalization}) and
(\ref{eq.normalization_symfunct}) is taken as the correlation time
($\tau_0 = 25$ ms) of the Ornstein-Uhlenbeck 
forcing of eq.~(\ref{eq.1}). For comparison we also show the
corresponding curves at $\alpha = 0.20, 0.51$ as blue circles and
red squares respectively, for which eq.~(\ref{eq.SSFT1}) holds.
The convergence to a master curve is verified, which means that
for a sufficiently strong forcing the thermal bath acts only as a
passive reservoir for the energy dissipation without providing any
important contribution to the energy injection into the
system. Evidently, the normalized asymmetry function for the
values $\alpha$ that verify the FT lie far from the master curve.
We point out that the transition to the limit $\alpha \gg 1$ is
rather continuous since intermediate regimes occur, as observed
for $\alpha = 1.84$. In this case neither the FT is satisfied as
shown previously in fig.~\ref{fig:FT_OPT} nor the master curve is
attained since the strength of thermal noise is still
comparable to that of the external noise.

The results for the normalized asymmetry function of the work done on the cantilever
by the external force are shown in fig.~\ref{fig:FTnormalized}(b). The curve corresponding
to the verification of the FT for $\alpha = 0.19$ is also plotted for comparison.
The convergence to a master curve is also checked as the value of $\alpha$ increases.
Indeed, when comparing our normalized experimental curves with the analytic expression
carried out by \cite{farago} for the asymmetry function of the work distribution on a Brownian particle
driven entirely by a Gaussian white noise
\begin{equation}
\label{asymptotic_asymmetry}
    \rho^{*}(w^{*}) = \left\{
    \begin{array}{ccc}
    4w^{*} & w^{*} <1/3\\
    \frac{7}{4}w^{*} + \frac{3}{2}-\frac{1}{4w^{*}} & w^{*} \ge 1/3
    \end{array} \right.,
\end{equation}
we  check that the assumption of the convergence of the energy injection process into the cantilever
to that of a Langevin dynamics for a harmonic oscillator entirely dominated by the external noise
is valid. Finite $\alpha$ corrections can be detected for large values of $w_{\tau}^{*}$ indicating
that the thermal bath still influences the energy injection into the cantilever.
This corrections seem to vanish as the system is driven farther from equilibrium,
as observed in fig.~\ref{fig:FTnormalized} for $\alpha = 22.10$.

Finally, we point out that the profile of the master curve strongly depends on the kind of
stochastic force: a Gaussian Ornstein-Uhlenbeck process in the first example and
a Gaussian white noise in the second one.
Non-Gaussian extensions of the external random force are expected to lead to striking
modification of the fluctuation relations in the limit $\alpha \gg 1$,
as recently investigated for an asymmetric Poissonian shot noise \cite{baule}.

\section{Conclusions}
We have studied the FT for the work fluctuations in two experimental 
systems in contact with a thermal bath and driven out of equilibrium  
by a stochastic force. The main result of our study is that the validity
of FT is controlled by the parameter $\alpha$. For small $\alpha \lesssim 1$
we have shown that the validity of the steady state FT
is a very robust result
regardless the details of the intrinsic dynamics of the
system (first and second order Langevin dynamics) and the
statistical properties of the forcing (white and colored 
Gaussian noise). Indeed these specific features vanish when the
integration of $w_{\tau}$ is performed for $\tau$ much larger than
the largest correlation time of the system.

In contrast for large $\alpha \gtrsim 1$, when the randomness of the
system becomes dominated by the external stochastic forcing, we have shown that FT
is violated. For $\alpha \gg 1$ the results at different driving amplitudes can
be set on a master curve by defining a suitable effective
temperature which is a function of $\alpha$. We have shown that
this master curve is system dependent.

\acknowledgments
%Insert here the text.

\end{document}